# Multilevel Quantum Particle as a Few Virtual Qubits Materialization

Alexander R. Kessel and Vladimir L. Ermakov

Physico-Technical Institute, Russian Academy of Sciences, Kazan 420029, Russia

*e-mail:* ermakov@sci.kcn.ru

**ABSTRACT**

A conception of virtual quantum information bit - virtual qubit - is introduced. It is shown by means of virtual qubit representation that four states of a single quantum particle is enough for implementation of full set of the gates, which is necessary for creation an arbitrary algorithm for a quantum computer.

The physical nature and mutual disposition of four working states is of no significance, if there are suitable selection rules for the particle interaction with the external electromagnetic field pulses.

## 1. INTRODUCTION

The search of new physical systems, capable to be a medium for implementation of the quantum computation, is one of the direction which promises appreciable progress in quantum computer realization. For this aim it is necessary to find an adequate quantum system and assign qubits to a subset of its stationary states. As the next step one should find the physical means which cause the transitions between these states, realizing quantum gates. Besides that it is necessary to determine how to create initial state and how to readout final state.

Spin 1/2 nuclear magnetic resonance is one of the examples of such physical systems[1,2]. Two possible spin 1/2 stationary states represent naturally one information bit (quantum bit or qubit). Thus generally used quantum computer model has the following correspondence: one qubit - one two level particle. The possibility to use the arbitrary physical origin quantum object possessing four or more discrete energy levels as a basis element for storing and processing information in a quantum computer is considered in the present paper. Main results have been published[3].

In standard quantum computer models the exchange interaction is used to realize two qubit gates. To implement the gates it is necessary to turn on the particle interaction for accurate time intervals and to turn it off when a gate should not operate. These time intervals are determined by interaction constant and may be longer than coherence time. In addition, to turn the interactions off the application of the complicated experimental schemes are necessary which can make the computation time very long.

In proposed model the gates are implemented on the single quantum particle and it is not necessary to spend time for the distant particle interactions have worked out. The logic gates in the proposed model are realized using external field short pulses. The pulse duration are defined by the field amplitude and are under full experimental control.

## 2. PHYSICAL SYSTEM

To be concrete let us consider four level discrete spectrum of a nucleus with spin 3/2 and quadruple moment, which is placed in a constant magnetic field and crystal electric field gradient.

### 2.1. The main Hamiltonian
The spin energy levels are defined by the Hamiltonian:

$$\mathbf{H}_0 = \mathbf{H}_Z + \mathbf{H}_Q, \tag{1}$$

where $\mathbf{H}_Z$ - Hamiltonian of interaction with external constant magnetic field, $\mathbf{H}_Q$ - Hamiltonian of electric quadrupole interactions. In particular case, when constant magnetic field is parallel to one of the electric field gradient principal axes, these operators take the form[4]

$$\mathbf{H}_Z = -\hbar\omega_0 \mathbf{I}_Z, \qquad \mathbf{H}_Q = \tfrac{1}{3}\hbar\omega_Q [3\mathbf{I}_z^2 - I(I+1) + \eta(\mathbf{I}_x^2 - \mathbf{I}_y^2)] \qquad (2)$$

where $\omega_Q$ - quadrupole interaction constant, $\omega_0$ - Zeeman frequency, $\eta$ - the electric field gradient asymmetry parameter ($|\eta|\leq 1$). In the spin 3/2 case the Hamiltonian (1),(2) eigenvalues are equal to

$$\hbar\varepsilon_{\pm 3/2} = \hbar\omega_Q (B_\mp \mp C); \qquad \hbar\varepsilon_{\pm 1/2} = \hbar\omega_Q (-B_\pm \pm C); \qquad (3)$$

where

$$B_\pm = \sqrt{[(1\pm 2C)^2 + \eta^2/3]}; \qquad C = \omega_0/\omega_Q;$$

The corresponding eigenfuctions are

$$|\Psi_{\mp 3/2}\rangle = cos(\alpha_\pm)|\chi_{\mp 3/2}\rangle + sin(\alpha_\pm)|\chi_{\pm 1/2}\rangle;$$
$$|\Psi_{\pm 1/2}\rangle = cos(\alpha_\pm)|\chi_{\pm 1/2}\rangle - sin(\alpha_\pm)|\chi_{\mp 3/2}\rangle; \qquad (4)$$

where $|\chi_m\rangle$ is the operator $\mathbf{I}_z$ eigenfunction corresponding to eigenvalue $m$ and

$$tg\,\alpha_\pm = (\sqrt{3})[B_\pm + (1\pm 2C)]/\eta.$$

To simplify the notation the indexes -3/2, -1/2, +1/2, +3/2 will be replaced by 1, 2, 3, 4 correspondingly. The above formulas (1)-(4) are valid at arbitrary ratio $R = \omega_0/\omega_Q$. To be concrete the case when the magnetic resonance spectrum is split by quadrupole interaction ($\omega_0 > \omega_Q$), at the condition that $\omega_Q$ is much grater than the spin level width is considered below. In this case the spin spectrum consists of the four well resolved resonance lines with energies $\varepsilon_1 > \varepsilon_2 > \varepsilon_3 > \varepsilon_4$.

## 2.2. The resonance radio frequency field interaction Hamiltonian

Neglecting the relaxation, the full nuclear spin Hamiltonian can be written as

$$\mathbf{H} = \mathbf{H}_0 + \mathbf{H}_{rf}, \qquad (5)$$

where $\mathbf{H}_{rf}$ is the operator of interaction with RF field. In the case when the RF field vector is parallel to Y axis this operator is

$$\mathbf{H}^Y_{rf}(t) = 2\hbar\gamma\, H_{rf} I_Y \cos(\Omega t + f), \qquad (6)$$

where $H_{rf}$, $f$ и $\Omega$ are the RF field amplitude, phase and frequency, $\gamma$ is the nucleus gyro magnetic ratio.

It will be needed later the operator $\mathbf{H}_{rf}(t)$ interaction representation form:

$$\mathbf{H}^*_{rf}(t) \equiv \mathbf{D}^{-1}(t-t_0)\, \mathbf{H}_{rf}(t)\, \mathbf{D}(t-t_0),$$
$$\mathbf{D}(t-t_0) = \exp[-i\mathbf{H}_0(t-t_0)/\hbar], \qquad (7)$$

The transformation operator $\mathbf{D}$ form corresponds to the case when the main Hamiltonian is time independent.

The representation of projective operators, $\mathbf{P}_{mn}$, for spins will be used to simplify notations and calculations. Each of these operators $\mathbf{P}_{mn}$ is 4x4 matrix with all zero elements except one $p_{mn} = 1$. Here $m,n = 1,2,3,4$ and thus one has the set of 16 operators. The spin components are expressed in the projective operator terms as

$$I_\alpha = \Sigma_{m,n} \langle \Psi_m | I_\alpha | \Psi_n \rangle \mathbf{P}_{mn}$$

The projective operators have very simple multiplication rules

$$\mathbf{P}_{kl}\mathbf{P}_{mn} = \delta_{lm}\mathbf{P}_{kn}, \qquad\qquad \mathbf{P}_{mn} = \mathbf{P}^+_{nm}$$

$$\mathbf{P}_{mn}|\Psi_k\rangle = \delta_{nk}|\Psi_m\rangle \tag{8}$$

In the projective operator representation the main Hamiltonian has the form

$$\mathbf{H}_0 = \Sigma_m \hbar\varepsilon_m \mathbf{P}_{mm}, \tag{9}$$

and the transformation operator is

$$\mathbf{D}(t-t_0) = \Sigma_m \mathbf{P}_{mm} \exp[-i\varepsilon_m(t-t_0)]. \tag{10}$$

The RF interaction Hamiltonian can be represented as

$$\mathbf{H}^*_{rf}(t) = \mathbf{\mathit{H}}_{rf,\,eff} + \Sigma_{kl} \mathbf{G}_{kl}\, exp[\,i(\varepsilon_k-\varepsilon_l)t] \tag{11}$$

when the RF frequency $\Omega$ coincides with one of the spin eigenfrequencies, for example with $\Omega_{mn} = |\varepsilon_m - \varepsilon_n|$. All operators in the equation (11) are of the same order of value: $|\mathbf{G}_{kl}| \sim |\mathbf{H}_{rf,\,eff}| \sim |\mathbf{H}_{rf}|$, and both $\mathbf{G}_{kl}$ and $\mathbf{H}_{rf,\,eff}$ are time independent. The last terms in (11), having amplitudes $|\mathbf{G}_{kl}|$, oscillate with eigenfrequencies $\Omega_{kl} \neq \Omega_{mn}$. Their influence on the spin evolution can be neglected in comparison with that of the constant term $|\mathbf{H}_{rf,eff}|$. This is due to the fact that in NMR experiments the RF pulse duration is usually much more than the cycle period. Thus the interaction (6) excites effectively resonance transitions between the energy levels with $\Omega = \Omega_{mn}$ only. That is why the interaction representation operator (6) is reduced over several cycles ($\Omega_{kl}\, t \gg 1$) to effective operator

$$\mathbf{H}^Y_{rf,\,eff} = i\,\hbar\gamma\, H_{rf} |\langle\Psi_m|\mathbf{\mathit{I}}_Y|\Psi_n\rangle|(\mathbf{P}_{mn}\, e^{if} - \mathbf{P}_{nm}\, e^{-if}). \tag{12}$$

It is necessary to point out that although in the basis $|\chi\rangle$ the operators $\mathbf{\mathit{I}}_X$, $\mathbf{\mathit{I}}_Y$ have matrix element selection rules $m-n = \pm 1$, in the basis $|\Psi\rangle$ these operators have non zero matrix elements with

m-n = ±1, ±2. (13)

**2.3. The spin system evolution under RF pulses**

The state vector evolution

$$|\Psi(t)\rangle = \mathbf{U}(t,t_0) |\Psi(t_0)\rangle, \quad (14)$$

of a physical system can be defined using unitary evolution operator[5]:

$$\mathbf{U}(t,t_0) = \mathbf{D}(t-t_0) \mathbf{V}(t,t_0), \quad (15)$$

$$\mathbf{V}(t,t_0) = \mathbf{T} \exp\left[-(i/\hbar) \int_{t_0}^{t} \mathbf{H}^*_{rf}(t') dt'\right],$$

$$\mathbf{H}^*_{rf}(t) = \mathbf{D}^{-1}(t-t_0) \mathbf{H}_{rf}(t) \mathbf{D}(t-t_0),$$

where $\mathbf{T}$ is chronological Dyson operator and $\mathbf{V}$ is the evolution operator in interaction representation.

At the condition, when the fast oscillating terms in $\mathbf{H}^*_{rf}(t)$ can be neglected, the chronological exponent transforms to ordinary one and the operator $\mathbf{V}(t,t_0)$ reduces to

$$\mathbf{V}_Y(t,t_0) \approx exp\left[-i/\hbar \int_{t_0}^{t} \mathbf{H}^Y_{rf,\,eff}\, dt'\right] = exp\left[(\varphi/2)(\mathbf{P}_{mn} e^{if} - \mathbf{P}_{nm} e^{-if})\right] \equiv \mathbf{V}_Y(\Omega_{mn}, \varphi_y).$$

where $\varphi_y = 2(t-t_0)\gamma H_{rf}|\langle\Psi_n|I_y|\Psi_m\rangle|$ and $\varepsilon_m > \varepsilon_n$.

Expanding the exponent and using the projective operator multiplication rules (8) one has

$$\mathbf{V}_Y(\Omega_{mn}, \varphi_y) = \mathbf{P}_{kk} + \mathbf{P}_{ll} + (\mathbf{P}_{nn} + \mathbf{P}_{mm})cos(\varphi_y/2) + (\mathbf{P}_{mn} e^{if} - \mathbf{P}_{nm} e^{-if}) sin(\varphi_y/2). \quad (16)$$

Here the indexes k,l ≠ m,n. The evolution operator expression $\mathbf{V}_X(\Omega_{mn}, \varphi_x)$ for the case when RF field is parallel to X axis can be obtained from Eq.(16) substituting the phase *f* by $(f - \pi/2)$. In the present paper all necessary quantum gates can be realized with phase $f=0$.

To implement quantum gates two frequency excitation will be necessary too. Below it will be shown that excitation of two transitions which have no common energy levels are enough for this aim. In these cases the evolution operator is a product of two commuting evolution operators which belongs to different transitions. For two excitation frequencies - $\Omega_{12}$ and $\Omega_{34}$ - the straightforward calculations give

$$\mathbf{V}_Y(\Omega_{12}, \varphi_Y; \Omega_{34}, \varphi'_Y) = (\mathbf{P}_{11} + \mathbf{P}_{22}) \cos(\varphi_Y/2) + (\mathbf{P}_{21} - \mathbf{P}_{12}) \sin(\varphi_Y/2) +$$
$$+ (\mathbf{P}_{33} + \mathbf{P}_{44}) \cos(\varphi'_Y/2) + (\mathbf{P}_{43} - \mathbf{P}_{34}) \sin(\varphi'_Y/2). \qquad (17)$$

For two excitation frequencies - $\Omega_{13}$ and $\Omega_{24}$ - the evolution operator is

$$\mathbf{V}_Y(\Omega_{13}, \varphi_Y; \Omega_{24}, \varphi'_Y) = (\mathbf{P}_{22} + \mathbf{P}_{44}) \cos(\varphi_Y/2) + (\mathbf{P}_{42} - \mathbf{P}_{24}) \sin(\varphi_Y/2) +$$
$$+ (\mathbf{P}_{33} + \mathbf{P}_{11}) \cos(\varphi'_Y/2) + (\mathbf{P}_{31} - \mathbf{P}_{13}) \sin(\varphi'_Y/2). \qquad (18)$$

### 3. THE VIRTUAL SPIN FORMALISM

In the adopted NMR computer standard model[1,2] the basis for quantum gates implementation consists of two *real* spins R=1/2 and S=1/2 with exchange interaction. The states of such a system are written usually in the quantum theory language as the abstract four dimensional space which is the direct product $\Gamma_R \otimes \Gamma_S$ of *real* spin two dimensional eigenstate spaces $\Gamma_R$ and $\Gamma_S$. To make clear the information aspect of suggested logic operations in our case it is useful to apply the inverse procedure: to represent *real* spin 3/2 four dimensional space $\Gamma_I$ as a direct product $\Gamma_R \otimes \Gamma_S$ of two abstract two dimensional spaces of the *virtual* spin R and S states. Then any four dimensional operator **P** can be expressed as a linear combination of direct product **R**⊗**S** components. There is the following isomorphic correspondence between the space $\Gamma_I$ basis $|\Psi_M\rangle$ and the direct product $\Gamma_R \otimes \Gamma_S$ basis $|\xi_m\rangle \otimes |\zeta_n\rangle$

$|\Psi_1> = |\xi_1>\otimes|\zeta_1> \equiv |11>, \qquad |\Psi_3> = |\xi_2>\otimes|\zeta_1> \equiv |01>,$

$|\Psi_2> = |\xi_1>\otimes|\zeta_2> \equiv |10>, \qquad |\Psi_4> = |\xi_2>\otimes|\zeta_2> \equiv |00>,$  (19)

where virtual spin indexes -1/2 and +1/2 are replaced with 1 and 2, correspondingly. Here |11> and so on are the notations, adopted in quantum information theory to represent the two qubit states. There are the following useful relations between the different spaces projective operators

$\mathbf{R}_{kl}\otimes\mathbf{S}_{mn}= \mathbf{P}_{2k-2+m,\ 2l-2+n}$ ;

$\mathbf{R}_{kl}\mathbf{R}_{mn} \otimes \mathbf{1}_S = \delta_{lm}\mathbf{R}_{kn} \otimes \mathbf{1}_S; \quad \mathbf{R}_{kl} \otimes \mathbf{1}_S |\xi_m>|\zeta_n> = \delta_{lm} |\xi_k>|\zeta_n>$ ;  (20)

$\mathbf{1}_R \otimes \mathbf{S}_{kl}\mathbf{S}_{mn} = \delta_{lm}\mathbf{1}_R \otimes \mathbf{S}_{kn}; \quad \mathbf{1}_R \otimes \mathbf{S}_{kl}|\xi_m>|\zeta_n> = \delta_{ln} |\xi_m>|\zeta_k>$ ;

$(\mathbf{R}_{kl}\otimes\mathbf{S}_{ab})(\mathbf{R}_{mn}\otimes\mathbf{S}_{cd})= \delta_{lm}\delta_{bc} \mathbf{R}_{kn} \otimes \mathbf{S}_{ad}$ ;

Here the operators $\mathbf{R}_{mn}$ and $\mathbf{S}_{mn}$ are the $\Gamma_R$ and $\Gamma_S$ space projective operators whereas $\mathbf{1}_R$ and $\mathbf{1}_S$ are the unit operators in the corresponding spaces. In particular the virtual spin components $\mathbf{R}_x$, $\mathbf{R}_y$, $\mathbf{R}_z$ and $\mathbf{S}_x$, $\mathbf{S}_y$, $\mathbf{S}_z$ are expressed as

$\mathbf{R}_x= (\mathbf{R}_{12}+\mathbf{R}_{21})/2; \qquad \mathbf{R}_y= i(\mathbf{R}_{12}-\mathbf{R}_{21})/2; \qquad \mathbf{R}_z=(\mathbf{R}_{11}-\mathbf{R}_{22})/2;$

$\mathbf{S}_x= (\mathbf{S}_{12}+\mathbf{S}_{21})/2; \qquad \mathbf{S}_y= i(\mathbf{S}_{12}-\mathbf{S}_{21})/2; \qquad \mathbf{S}_z=(\mathbf{S}_{11}-\mathbf{S}_{22})/2.$

It is necessary to point out the fact that in four dimensional space $\Gamma_I$ two qubits can be defined in another way. One can consider that the upper two levels compose the first qubit whereas the bottom pair of levels composes the second qubit. Such two level system definition is commonly used in magnetic resonance under the name «*fictitious* (or *effective*) spin 1/2 formalism». In this case the space $\Gamma_I$ is a direct sum $\Gamma_R\oplus\Gamma_S$ of *fictitious* spin spaces. Using such an approach one can easily implement single qubit rotations, but the two qubit «controlled NOT» gate implementation will face with some problems. Besides that the introduced virtual spins correspond very well to quantum information language.

# 4. INITIAL STATE PREPARATION

The state |00> is the initial input state for present days quantum algorithms. To emulate an abstract quantum computer by means of NMR experiments it is necessary to take into account the following peculiarities of NMR quantum computing. In NMR information processing the macroscopic number of identical quantum processors - molecules are used in parallel and the output is a sum of these molecules' signals. The density matrix formalism is the adequate language for NMR experiments. The density matrix $\sigma_{init} = const\, \mathbf{P}_{44}$ is the equivalent of the |00> state.

The macroscopic sample of identical nuclei ensemble has energy levels with population distributed according to Boltzman law

$$\rho_{eq} = Z^{-1} \exp(-\beta \mathbf{H}), \qquad Z = Sp[\exp(-\beta \mathbf{H})], \quad \beta = 1/\kappa T, \qquad (21)$$

In the usual NMR experiments $|\beta \mathbf{H}| \ll 1$ (normally $\sim 10^{-5} - 10^{-6}$). At such conditions the input state for NMR quantum computer is given by matrix

$$\rho_{eq} = Z^{-1}[\mathbf{1_I} + \Sigma_m \lambda_m \mathbf{P}_{mm}]$$
$$\mathbf{1_I} = \Sigma_m \mathbf{P}_{mm}; \qquad \lambda_m = \hbar\varepsilon_m/kT; \qquad (22)$$

It is possible to prepare $\rho_{eq}$ in the form $\sigma_{init}$ directly by cooling spin system till ultra low temperature. It will create not only tremendous technological difficulties but also will slow down the computation speed. Below it will be shown how in real NMR experiments to get the density matrix

$$\rho_{init} = const\,[\mathbf{1_I} + const\, \mathbf{P}_{44}] \qquad (23)$$

as an input for quantum algorithms. In comparison with $\sigma_{init}$ there is an additional term in $\rho_{init}$ which is proportional to unity matrix $\mathbf{1_I}$. But it does not contribute to observable signal being invariant under unitary transformations. Therefore any pulse sequences, including the computational ones, do not affect the matrix $\mathbf{1}_I$ and that is why the initial matrixes $\rho_{init}$ and $\sigma_{init}$ give the same output result.

The space[2] or temporal[6] averaging procedure can be used to transform the density matrix $\rho_{eq}$ to the desirable form $\rho_{init}$. Let us to consider temporal averaging procedure for multiqubit spin, leaving other methods for future developments.

The following procedure, which goes back to the paper[6] ideas, can be suggested. Let the desirable computation consists of the fulfilling the transformation $\mathbf{V}_{comp}$ with the initial state $\rho_{init}$, whereas spin system has an equilibrium density matrix of the form (22). It can be shown that the average

$$(1/3)[\mathbf{V}_{comp}\rho_{eq}\mathbf{V}^{\dagger}_{comp} + \mathbf{V}_{comp}\mathbf{V}_1\rho_{eq}\mathbf{V}^{\dagger}_1\mathbf{V}^{\dagger}_{comp} + \mathbf{V}_{comp}\mathbf{V}_2\rho_{eq}\mathbf{V}^{\dagger}_2\mathbf{V}^{\dagger}_{comp}] =$$
$$= (1/3)\mathbf{V}_{comp}(\rho_{eq} + \mathbf{V}_1\rho_{eq}\mathbf{V}^{\dagger}_1 + \mathbf{V}_2\rho_{eq}\mathbf{V}^{\dagger}_2)\mathbf{V}^{\dagger}_{comp}$$

of three computations with the same propagator $\mathbf{V}_{comp}$, all of which start from three different input states $\rho_{eq}$, $\mathbf{V}_1\rho_{eq}\mathbf{V}^{\dagger}_1$ and $\mathbf{V}_2\rho_{eq}\mathbf{V}^{\dagger}_2$, is equivalent to transformation $\mathbf{V}_{comp}$ of the state $\rho_{init}$:

$$\mathbf{V}_{comp}\rho_{init}\mathbf{V}^{\dagger}_{comp} \qquad (24)$$

To prove this statement we choose the transformations $\mathbf{V}_1$ and $\mathbf{V}_2$ in the form of two successive single frequency pulse propagators

$$\mathbf{V}_1 = \mathbf{V}_Y(\Omega_{12},\pi)\mathbf{V}_Y(\Omega_{23},\pi) = \mathbf{P}_{44} + \mathbf{P}_{21} + \mathbf{P}_{13} + \mathbf{P}_{32},$$
$$\mathbf{V}_2 = \mathbf{V}_Y(\Omega_{23},\pi)\mathbf{V}_Y(\Omega_{12},\pi) = \mathbf{P}_{44} - \mathbf{P}_{12} + \mathbf{P}_{31} - \mathbf{P}_{23} \qquad (25)$$

Using equations (16a), (16b) and multiplication rules (8) we get

$$(1/3)[\rho_{eq} + \mathbf{V}_1\rho_{eq}\mathbf{V}^{\dagger}_1 + \mathbf{V}_2\rho_{eq}\mathbf{V}^{\dagger}_2] = Z[\alpha\mathbf{1} + \beta\mathbf{P}_{44}] \propto \rho_{init}, \qquad (26)$$

$\alpha = 1+(1/3)[\lambda_1 +\lambda_2 + \lambda_3]$,    $\beta = \lambda_4 - (1/3)[\lambda_1 +\lambda_2 + \lambda_3]$.

This proves the possibility to start with $\rho_{eq}$ instead of $\rho_{init}$ using suitable pulse sequences.

## 5. ONE QUBIT ROTATION

The operator $\mathbf{V}_Y(\Omega_{12}, \varphi_Y; \Omega_{34}, \varphi'_Y)$, Eq. (17), at the condition $\varphi_Y = \varphi'_Y = \varphi$ is equal to

$$\mathbf{V}_Y(\Omega_{12}, \varphi; \Omega_{34}, \varphi) = (\mathbf{P}_{11}+\mathbf{P}_{22}+\mathbf{P}_{33}+\mathbf{P}_{44})cos(\varphi/2)+(\mathbf{P}_{21}-\mathbf{P}_{12}+\mathbf{P}_{43}-\mathbf{P}_{34})sin(\varphi/2). \qquad (27)$$

Expressing the operators $\mathbf{P}$ in terms of operators $\mathbf{R}$ and $\mathbf{S}$, Eq. (20), it can be shown that the operator

$$\mathbf{V}_Y(\Omega_{12}, \varphi; \Omega_{34}, \varphi) = (\mathbf{R}_{11} + \mathbf{R}_{22})\otimes[(\mathbf{S}_{11} + \mathbf{S}_{22})cos(\varphi/2) + (\mathbf{S}_{21} - \mathbf{S}_{12})sin(\varphi/2)] =$$
$$= exp\{-i\varphi\, \mathbf{1}_R\otimes\mathbf{S}_y\} \qquad (28)$$

is equivalent to operator (27). The last equality in the right hand side of (28) means that the transformation $\mathbf{V}_Y(\Omega_{12}, \varphi; \Omega_{34}, \varphi)$ is a rotation through the angle $\varphi$ around Y axis in the space $\Gamma_S$ keeping the space $\Gamma_R$ invariant.

Using the same arguments it can be shown that the operator $\mathbf{V}_Y(\Omega_{13}, \varphi_Y; \Omega_{24}, \varphi'_Y)$, Eq. (18), at $\varphi_Y=\varphi'_Y\equiv\varphi$ is equal to operator

$$\mathbf{V}_Y(\Omega_{13}, \varphi; \Omega_{24}, \varphi) = [(\mathbf{R}_{11}+\mathbf{R}_{22})cos(\varphi/2)+(\mathbf{R}_{21}-\mathbf{R}_{12})sin(\varphi/2)]\otimes(\mathbf{S}_{11}+\mathbf{S}_{22}) =$$
$$= exp\{-i\varphi\, \mathbf{R}_y\otimes\mathbf{1}_S\}. \qquad (29)$$

The last equality in the right hand side of (29) means that the transformation $\mathbf{V}_Y(\Omega_{13}, \varphi; \Omega_{24}, \varphi)$ is a rotation through the angle $\varphi$ around Y axis in the space $\Gamma_R$ keeping the space $\Gamma_S$ invariant.

# 6. «CONTROLLED NOT» GATE - CNOT

The transformation $\mathbf{V}_y(\Omega_{12}, \varphi_y)$, Eq. (16), which corresponds to energy levels $\hbar\varepsilon_1$ and $\hbar\varepsilon_2$ and is defined as

$$\mathbf{V}_y(\Omega_{12}, \pi) = [\mathbf{P}_{33} + \mathbf{P}_{44}] + [\mathbf{P}_{21} - \mathbf{P}_{12}], \tag{30}$$

at $\varphi_y = \pi$, implements two qubits operation «controlled NOT» - CNOT: NOT operation at spin S if R is in state $|\xi_1\rangle$ and leaves S in its current state if R is in state $|\xi_2\rangle$. Straightforward calculation using (30) gives

$\mathbf{V}_y(\Omega_{12}, \pi) |\Psi_1\rangle \equiv |\Psi_2\rangle,$   $\mathbf{V}_y(\Omega_{12}, \pi) |\Psi_2\rangle \equiv -|\Psi_1\rangle,$

$\mathbf{V}_y(\Omega_{12}, \pi) |\Psi_3\rangle \equiv |\Psi_3\rangle,$   $\mathbf{V}_y(\Omega_{12}, \pi) |\Psi_4\rangle \equiv |\Psi_4\rangle,$

or in information notations

$\mathbf{V}_y(\Omega_{12}, \pi) |11\rangle = |10\rangle,$   $\mathbf{V}_y(\Omega_{12}, \pi) |10\rangle = -|11\rangle,$

$\mathbf{V}_y(\Omega_{12}, \pi) |01\rangle = |01\rangle,$   $\mathbf{V}_y(\Omega_{12}, \pi) |00\rangle = |00\rangle.$

It is the truth table for CNOT operation defined above neglecting phase factor $exp(i\pi)$. The evolution operator $\mathbf{V}_y(\Omega_{12}, \pi)$ in the basis $\Gamma_R \otimes \Gamma_S$ projective operators representation can be written as

$$\mathbf{V}_y(\Omega_{12}, \pi) = \mathbf{R}_{11} \otimes (\mathbf{S}_{21} - \mathbf{S}_{12}) + \mathbf{R}_{22} \otimes \mathbf{1}_S \tag{31}$$

Using the same arguments it can be shown that transition transformation $\mathbf{V}_y(\Omega_{13}, \varphi_y)$ between $\hbar\varepsilon_1$ and $\hbar\varepsilon_3$ states at $\varphi_y = \pi$

$$\mathbf{V}_y(\Omega_{13}, \pi) = [\mathbf{P}_{22} + \mathbf{P}_{44}] + [\mathbf{P}_{31} - \mathbf{P}_{13}], \tag{32}$$

implements CNOT operation in which the spin S state controls the spin R state. Operator (32) gives the truth table

$\mathbf{V}_y(\Omega_{13}, \pi) |\Psi_1\rangle \equiv |\Psi_3\rangle,$ $\qquad \mathbf{V}_y(\Omega_{13}, \pi) |\Psi_2\rangle \equiv |\Psi_2\rangle,$

$\mathbf{V}_y(\Omega_{13}, \pi) |\Psi_3\rangle \equiv -|\Psi_1\rangle,$ $\qquad \mathbf{V}_y(\Omega_{13}, \pi) |\Psi_4\rangle \equiv |\Psi_4\rangle.$

which in the information notations has the form

$\mathbf{V}_y(\Omega_{13}, \pi) |11\rangle = |01\rangle,$ $\qquad \mathbf{V}_y(\Omega_{13}, \pi) |10\rangle = |10\rangle,$

$\mathbf{V}_y(\Omega_{13}, \pi) |01\rangle = -|11\rangle,$ $\qquad \mathbf{V}_y(\Omega_{13}, \pi) |00\rangle = |00\rangle.$

Using the basis $\Gamma_R \otimes \Gamma_S$ projective operators the evolution operator $\mathbf{V}_y(\Omega_{13}, \pi)$ can be written as

$$\mathbf{V}_y(\Omega_{13}, \pi) = \mathbf{1}_R \otimes \mathbf{S}_{22} + (\mathbf{R}_{21} - \mathbf{R}_{12}) \otimes \mathbf{S}_{11}. \tag{33}$$

Thus the spin I=3/2 pulse excitation corresponding to the evolution operators $\mathbf{V}_y(\Omega_{12},\pi)$ and $\mathbf{V}_y(\Omega_{13},\pi)$ implements the quantum CNOT gates of two kinds - when spin R controls the spin S dynamics and vice versa.

## 7. READOUT

To obtain results of computation it is necessary to have possibility to read out the density matrix final state $\rho_{out}$. The NMR methods allows to get all matrix elements using so called «state tomography»[7]. This method is realized by means of complicated pulse sequences. For other physical systems the development of adequate means is necessary.

## 8. CONCLUDING REMARKS

The presented formalism is applicable to quantum system of arbitrary physical nature. One can use the four appropriate energy levels of greater nuclear spins, electron paramagnetic resonance spectra with effective spin $S^* \geq 3/2$, optical levels and so on. Only the resonance frequency values and the operators matrix elements will be changed in different physical realizations. For example, the nuclear spin I=3/2 quadrupole resonance spectrum being split by interaction with constant magnetic field ($\omega_0 < \omega_Q$) can be used. This case differs from the one considered in the paper only in mutual energy levels arrangement whereas the equations (1)-(10) for operators and eigenfunctions remain the same.

The virtual spins representation provides a new way to implement three qubit gates. In future such gates can be useful for development the compact algorithms or for other purposes. But its implementation using two level quantum systems requires three body interactions which are absent in nature. One can notice that spin 7/2 Hilbert space $\Gamma_I$ can be considered as three virtual spin 1/2 Hilbert spaces direct product. Therefore the Hilbert space of one such particle can contain three qubits. Then, finding suitable pulse sequences that act onto eight dimensional spin 7/2 system, it is possible to implement the gates, for realization of which three individual spins 1/2 interaction is necessary. In[8] another solution of this problem was found and it was shown that special two qubit gate set based on two body physical interactions is enough to realized any arbitrary quantum algorithm.